\documentclass[12pt]{article}

\usepackage{amsmath}
\usepackage{amssymb}
\usepackage{graphicx}
\newcommand{\be}{\begin{equation}}
\newcommand{\ee}{\end{equation}}
\newcommand{\cpn}{CP$^{N-1}$}
\newcommand{\p}{\phantom{0}}

\usepackage{a4}

\begin{document}
\begin{flushright}
\small{CERN-PH-TH/2011-023}
\end{flushright}
\vspace{1cm}

\begin{center}
{\LARGE Testing trivializing maps in the Hybrid Monte Carlo algorithm}\\
\vspace{1cm}

{\large Georg P.~Engel}\\
{Institut f\"ur Physik, FB Theoretische Physik, \\Universit\"at Graz,
A-8010 Graz, Austria}\\[2ex]
{\large Stefan Schaefer}\\
{CERN, Physics Department, 1211 Geneva 23, Switzerland}
\end{center}
\vspace{0.5cm}

\date{}

\begin{abstract}
We test a recent proposal to use approximate trivializing maps in a
field theory to speed up Hybrid Monte Carlo simulations. Simulating the
\cpn model, we find a small improvement with the leading order
transformation, which is however compensated by the additional
computational overhead.  The scaling of the algorithm towards
the continuum is not changed. In particular, the effect of the
topological modes on the autocorrelation times is studied. 
\end{abstract}

\section{Introduction}

In the simulation of statistical models, many Monte Carlo methods experience a
significant increase in effort when approaching a continuous phase transition
of the theory.  This phenomenon is called critical slowing down and depends
strongly on the nature of the underlying theory and the algorithm used. For
some models, algorithms have been found which completely eliminate this slowing
down or even lead to a speed-up when the critical line is approached. 
A particular type of critical slowing down is associated with the
topological modes of the theories. In QCD, e.g., the topological charge
of the gauge configuration is known to be particularly problematic with
both single link updates \cite{DelDebbio:2002xa} and algorithms based
on molecular dynamics \cite{Schaefer:2010hu}.  

Recently, a possible solution to the problem has been
proposed \cite{Luscher:2009eq}, for which field transformations given by flow
equations are introduced.  Exactly integrating the flow equations, the
theory becomes  trivial and therefore also trivial to simulate.  The
problem is that the differential equations generating the flow are not
exactly known, however, the first terms of a power series of the
corresponding kernel can easily be constructed.  At this point, it is
unclear whether it is sufficient to just know the first order of the
differential equations to sufficiently mitigate the problem. Also the
accuracy to which the differential equations have to be integrated is
unknown.

In this letter, we therefore put this proposal to a test. Since it should work
for any field theory, we simulate the \cpn model for $N=10$ using the Hybrid
 Monte Carlo (HMC) algorithm \cite{Duane:1987de}, which is an integral part of the program
laid out in Ref.~\cite{Luscher:2009eq}. We choose this model, because it shares
some similarities with QCD like asymptotic freedom and confinement and, like in
QCD, the topological modes show a much more severe critical slowing down than
other observables \cite{DelDebbio:2004xh}.  Most importantly, the accuracy
achievable in this two dimensional model is much higher than in QCD and the
reduced cost also allows for a better mapping of the rather high dimensional
parameter space of the problem. 

In Sect.~\ref{sec:setup} we give the details of the model and how to set up the
HMC algorithm for it.  After that, we discuss the trivializing map and its
construction to leading order in Sec.~\ref{sec:TrivMap}.  Then we give the parameters of the simulation
which enables us to put the approximate trivializing map to a test whose
results we give in Sect.~\ref{sec:sim}.
In the definition of the action, the observables and as a point of
reference for the main quantities, we follow the paper by Campostrini,
Rossi and Vicari~\cite{Campostrini:235846}.

\section{The \cpn model  and Hybrid Monte Carlo\label{sec:setup}}

We immediately give the model on a square lattice with lattice spacing $a$ and
sites $n=a(n_1, n_2)$, $n_1$ and $n_2$ integer numbers.  The fields living on
the sites are complex $N$ component unit vectors $z_n$ connected by U(1) link
variables $\lambda_{n,\mu}$, which are represented by complex numbers on the
unit circle. The action is then
given by \cite{Rabinovici:1980dn}
\begin{equation}
S[z,\lambda]	
= -N \beta \sum_n \sum_{\mu=1}^2\left(z_{n+\hat\mu}^\dagger z_n\lambda_{n,\mu} 
   + z_{n}^\dagger z_{n+\hat\mu}\lambda_{n,\mu}^*-2 \right) \;.
   \label{eq:act1}
\end{equation}
The gauge fields $\lambda_{n,\mu}$ can be integrated out analytically,
thus this lattice action is expected to lie within the universality class of the \cpn model.

For the Hybrid Monte Carlo algorithm as well as for the field
transformation we will need derivatives with respect to the field
degrees of freedom.  Therefore it is convenient to treat the fields as $real$
$2N$ component fields $x_n$, which live on the unit sphere in
$\mathbb{R}^{2N}$
\begin{equation}
x_{n,2i} = \mathrm{Re}{(z_{n,i})} \;, \;\; 
x_{n,2i+1} = \mathrm{Im}{(z_{n,i})} \;\;\;, \; i=0,\dots,N-1 \;.
\end{equation}
In the following, we will make use of either $z$ or $x$, such that
 each formula appears in its most simple form.

The $U(1)$ fields $\lambda$ are naturally parametrized by  the angle
$\phi \in [0,2\pi)$ with $\lambda=e^{i\phi}$. Its action on the vectors
$x$ is then given by SO(2) $2N\times2N$ matrices
$\Lambda_{n,\mu}(\phi)$, which are zero everywhere but on the diagonal
$2\times2$ blocks
\[
\begin{pmatrix} 
\Lambda_{2i,2i} & \Lambda_{2i,2i+1} \\
\Lambda_{2i+1,2i} & \Lambda_{2i+1,2i+1}
\end{pmatrix}
=
\begin{pmatrix} 
\cos \phi &  -\sin \phi \\
\sin \phi & \cos \phi
\end{pmatrix}
\;\;, \; i=0,\dots,N-1 \;.
\]
With these definitions, we can write the action as function of real variables only
\begin{equation}
S[x,\phi]	= -N \beta \sum_n \left(x_n^T J_n -4 \right)  \;,
\label{eq:act}
\end{equation}
where we introduced the ``spin sum'' of gauge-transported nearest neighbors
\begin{eqnarray}
J_n	&=& \sum_{\mu=\pm1}^{\pm2} \Lambda_{n,\mu}^T x_{n+\hat\mu}  
\;,  \;\;\; \text{with} \;\;\Lambda_{n,-\mu}:=\Lambda_{n-\hat\mu,\mu}^T  \;.
\end{eqnarray}
The partition function can then be easily written by embedding the unit vectors
$x_n$ into $\mathbb{R}^{2N}$.
\subsection{Observables}
We will focus our study on a few, central observables. Mainly the energy
density $E=S/(N\beta V)$, the magnetic susceptibility $\chi_M$ and the
correlation length $\xi_G$. The latter two are constructed from the two
point function in momentum space 
\begin{equation}
\widetilde G_P({k})= \frac{1}{V} \sum_{n,m} \langle \mathrm{tr} P_{n}
P_{m} \rangle_\mathrm{conn} \exp \left (\frac{2\pi i}{L} ({
n-m}) \cdot {k} \right ) 
\end{equation}
with $ P_{n} = z_n z_n^\dagger  $.
Using these definitions, the two remaining observables are then
\begin{equation}
\chi_M=\widetilde{G}_P(0,0) \ ;  \qquad \xi_G^2=\frac{1}{4\sin^2
\frac{\pi}{L}} \left ( \frac{\widetilde{G}_P(0,0)}{\widetilde{G}_P(0,1)}-1
\right ) \ . 
\end{equation}
The topological charge density $q_n$ is given by the sum over the angles
between the spins around a plaquette 
\[
q_n=\frac{1}{4\pi}\epsilon_{\mu \nu} 
(\theta_{n,\mu}+\theta_{n+\hat\mu,\nu}-\theta_{n+\hat\nu,\mu}-\theta_{n,\nu})
\; \mathrm{mod} \; 1 \; ; \;\; -\frac{1}{2} < q_n \leq \frac{1}{2}
\]
with $\theta_{n,\mu}=\mathrm{arg}(z_n^\dagger z_{n+\hat\mu})$. 
The topological charge is the volume integral of this quantity
$Q=\sum_nq_n$.

\subsection{Hybrid Monte Carlo\label{sec:hmc}}
In Hybrid Monte Carlo, the fields $x$ and $\phi$ are updated by introducing
conjugate momenta $\pi$ and $\omega$ and solving classical equations
of motion associated with the Hamiltonian 
\[
H[\pi, \omega,x,\phi]=\frac{1}{2}\sum_n (\pi_n)^2
+\frac{1}{2}\sum_{n,\mu} (\omega_{n,\mu})^2 +S[x,\phi] \;.
\]
The momenta live in the tangent spaces of the respective field manifolds
and therefore $\pi_n \in \mathbb{R}^{2N}$ with ${\pi_n}\cdot x_n=0$ and
$\omega_{n,\mu} \in \mathbb{R}$ without further conditions, because the
manifold is flat. The Hamilton equations of motion are then
\begin{align}
\dot x_n   &= \pi_n \; ,                      & \dot \phi_{n,\mu}
&=\omega_{n,\mu} \; ,\\
\dot \pi_n &= -\widetilde \nabla_{x_n} S[x,\phi]\; , & \dot
\omega_{n,\mu} &= -\partial_{\phi_{n,\mu}} S[x,\phi] \; .
\label{eq:eqm}
\end{align}
A natural derivative $\widetilde\partial_x^i$ of a function $f(x)$ defined on the unit sphere
is the projection of the ordinary gradient $\nabla_x$ in $\mathbb{R}^{2N}$ onto
 the tangent space of the sphere
\begin{equation}
\widetilde{\partial}_{x}^i f\left(x\right) =
\left[\left(\mathbf1-xx^T\right) \nabla_{x} f(x) \right]_i \ . \label{eq:deriv}
\end{equation}
This corresponds to continuing the function $f(x)$ to the full $\mathbb{R}^{2N}$ via
$\tilde f(x) = f(x/|x|)$ and then taking ordinary derivatives.
The forces in the equations of motion~(\ref{eq:eqm}) then read for the
action given in Eq.~(\ref{eq:act})
\begin{equation}
F^x_n	= - \tilde{\nabla}_{x_n} S[x,\phi] 
        = 2N\beta\left(\mathbf1-x_n x_n^T\right) J_n =: 2N\beta p_n \;,
\label{forcex}
\end{equation}
where we have defined $p_n$ as the projection of the spin sum $J_n$ to the tangent space at $x_n$.
The forces $F^\phi_{n,\mu}$ for the conjugate momenta $\omega_{n,\mu}$ are
\begin{equation}
\label{forcephi}
F^\phi_{n,\mu}	= -{\partial}_{\phi_n} S[x,\psi] = -2N\beta x_n^T \Gamma \Lambda_{n,\mu}^T x_{n+\hat\mu} 
\quad\text{with}\quad 
\Gamma =
\begin{pmatrix} 
0 &  -1 \\
1 & 0
\end{pmatrix}\;.
\end{equation}
The $\Gamma$ is the translation of the imaginary unit $i$ to the
language of the $2$ component real vectors. 

In the numerical simulations, we use a leap-frog integration scheme
with a single time scale. For this, we need finite step size updates
of the fields to numerically solve Eqs.~(\ref{eq:eqm}). The only non
trivial part is the update of the field $x_n$ with momentum $\pi_n$, 
because also the updated variables have to fulfill the constraints
$|x_n'|=1$ and $x_n'\cdot \pi'_n=0$. For an infinitesimal step of size
$\epsilon$, we therefore use the map $\Phi_\epsilon$
\begin{equation}
\begin{pmatrix}
x'\\ \pi'
\end{pmatrix}
=
\Phi_{\epsilon}(x,\pi)=
\begin{pmatrix}
\cos \alpha & \frac{1}{|\pi|} \sin \alpha  \\
-|\pi| \sin \alpha & \cos \alpha
\end{pmatrix}
\begin{pmatrix}
x \\ \pi
\end{pmatrix}
\quad\text{with}\quad
\alpha=\epsilon|\pi|  \ .
\end{equation}
It corresponds to the exact solution of the equation of motion in the absence of 
the forces $F$ but subject to the constraint $|x|=1$.

\section{Trivializing map in the \cpn model\label{sec:TrivMap}}

The goal of the field trivialization is to find a map $\mathcal{F}$ in
field space such that the Jacobian $\mathcal{J}$ of the transformation 
compensates the action. For the partition
function Eq.~(\ref{eq:act1}) this would mean a transformation
$(x',\phi')=\mathcal{F}^{-1}(x,\phi)$ such that
\begin{equation}
Z=\int [d x][d \phi] e^{-S[x,\phi]}
=\int [d x'][d \phi']e^{-S[\mathcal{F}(x',\phi')]+\log \det
\mathcal{J}[x',\phi']}
\label{eq:trans}
\end{equation}
with the exponent equal to a constant. Since 
in this case all configurations in the new variables are equally likely,
the molecular dynamics evolution of the HMC algorithm would not experience
any forces and be very efficient. But also in the situation that one
can only find an approximation to the exact trivializing  map $\mathcal{F}$,
one can expect a significant gain in the performance of the algorithm.

In our simulations, the forces associated to the U$(1)$ field $\lambda$
are much smaller than those associated to the spin variables $z$. For
all considered $\beta=0.7,\dots,1.0$ in CP$^9$, we found the ratio of the average forces to be
about ten and also for the maximal force a factor of almost four.
We therefore perform the field transformation only on the $x$, leaving
the $\phi$ untouched. In the remaining part of this section, we go
through the major steps of the computation, following the lines of
Ref.~\cite{Luscher:2009eq}.

The trivializing map $\mathcal{F}$ can be obtained by integrating a flow $T$ from $t=0$ to $t=1$.
Note, however, that integration to some $t_T\leq1$ will probably be a better choice for only approximate trivializing flows.
A possible ansatz for the flow $T$ is to take a gradient of an action $\tilde{S}$
\begin{equation}
\dot x_n^i(t)=-\tilde \partial^i_n  \tilde{S}[t,x(t)]\equiv T_n^i[t,x(t)] \ . \label{eq:flow} 
\end{equation}
The action $\tilde S$ can be expanded in a power series in $t$
\[
\tilde S[t,x]=\sum_{k=0}^{\infty} t^k \tilde S^{(k)}[x]
\]
for which the leading order term can be constructed easily. 
One result of Ref.~\cite{Luscher:2009eq} is that the leading term of $\tilde S$ fulfills 
\[
-\sum_n {\tilde \partial}^i_n {\tilde\partial}^i_n \tilde S^{(0)}=S+C\;.
\]
Using the derivative defined in Eq.~(\ref{eq:deriv}), it is easy to show that on the unit
sphere 
\[
-\tilde\partial^i\tilde \partial^i f(x)= (2N-1) x\cdot\nabla f(x)-\mathrm{tr}\left [ (1-x x^T) H_f(x) \right ]
\]
with the Hessian $H_f(x)_{ij}=\partial_i\partial_jf(x)$. 
This immediately leads to 
\begin{equation}
\tilde{S}^{(0)}		= \frac{1}{2(2N-1)} S \;.
\end{equation}
The corresponding leading order trivializing flow from
Eq.~(\ref{eq:flow}) reads
\begin{equation}
\label{eq:TrivMapLO}
T_n^{(0)} = \frac{2N\beta}{2N-1} p_n \;.
\end{equation}
As next step, we need a numerical integration scheme for the flow
equation (\ref{eq:flow}).
Following \cite{Luscher:2009eq}, we use an Euler integrator 
in which each step
is similar to the finite step-size update discussed in
Sec.~\ref{sec:hmc}
\begin{equation}
\label{Euler}
{x}_n(t+\epsilon_s)= \cos\alpha_nx(t)  + \sin\alpha_n \frac{T_n(t)}{|T_n(t)|}
\quad\text{with} \;\; \alpha_n	=\epsilon_s|T_n| \;.
\end{equation}
This integrator has $\mathcal{O}(\epsilon_s)$ errors when integrating to
a fixed $t$, but as we see below, the method does not suffer
significantly from these inaccuracies.  To get the action in the
transformed variables,  the determinant of the  Jacobian of the field
transformation has to be computed, see Eq.~(\ref{eq:trans}). Since this is
too complicated if all spins are changed at once, we follow
Ref.~\cite{Luscher:2009eq} again and transform one spin at a time,
sweeping through the lattice. This sweep is done for each step in
$\epsilon_s$ of the Euler integrator.

For the transformation given in Eq.~(\ref{Euler}), the determinant can be easily computed.
In the language of Eq.~(\ref{eq:trans}), for a single step the primed
quantities are at $t$, whereas the unprimed quantities at $t+\epsilon_s$.
This transformation only changes the angle $\theta$ between the neighbor sum $J_n$
and the transformed spin $x_n$, leaving the components perpendicular
to this plane untouched. Specifically, it is changed to
\[
\theta=\theta'-\alpha=\theta'-\frac{\epsilon_sN\beta}{2N-1} |J_n| \sin \theta' \;.
\]
Since the integration measure for the angular component of
 spherical coordinates in the $\mathbb{R}^{2N}$
is $(\sin\theta)^{2N-3} d \cos \theta$ one easily obtains
the Jacobi determinant as
\begin{equation}
\det \mathcal{J}_n= \left(1-\frac{\epsilon_sN\beta}{2N-1} J_n^Tx_n' \right) \left(\cos\alpha-\frac{J_n^T x_n'}{|p_n'|}\sin\alpha \right)^{2N-2} \;.
\end{equation}
The forces corresponding to the action constructed from the smoothed
fields $x(t)$, have to be computed using the chain rule. Since this is a
standard procedure, we do not describe it here in detail.

\section{Details of the simulation\label{sec:sim}}

To our knowledge, this investigation is the first using the HMC
algorithm to simulate the \cpn model. It is certainly not the algorithm
of choice for  this theory, but our objective is the study of
the improvement in the algorithm brought by the leading order
trivialization.  For comparison with the literature, we relied heavily
on Ref.~\cite{Campostrini:235846}, from which we reproduced several
observables for all parameter sets we looked at. 
Since there is no prior experience with the HMC in this model,
we have to first study it in the normal variables and can then assess 
the change that is brought by the field transformation.
We use the acronym THMC for the HMC with field transformation in the following.

The figure of merit is the autocorrelation time of interesting
observables in units of the molecular dynamics time. It is defined
via the autocorrelation function
\[
\Gamma_{A}(t)=\langle (A(s+t) - \bar A) (A(s) -\bar A) \rangle  \; ,
\]
where the average is over independent realizations of the Markov chain
and $\bar A$  is the expectation value of the observable $A$. The
integrated autocorrelation time is then 
\begin{equation}
\tau_\mathrm{int}(A) = \frac{1}{2} + \sum_{t=1}^\infty
\frac{\Gamma_A(t)}{\Gamma_A(0)} \ .
\label{eq:tauint}
\end{equation}
Since Monte Carlo histories are never infinitely long, the sum in
Eq.~(\ref{eq:tauint}) has to be truncated at some window $W$. For its
choice, one has to balance
the statistical error, which increases with larger windows, with the
systematic error from neglecting the contributions beyond $W$. For this
purpose, 
we use the software described in Ref.~\cite{Wolff:2003sm}. In most cases,
its automatic criterion turned out to be sufficient, however, due to our
high statistics data, we sometimes had to increase the parameter $S$,
which influences the relative size between the window $W$ chosen and
$\tau_\mathrm{int}(W)$. Also because of the high statistics, we are
confident that the systematic error due to slow modes is under
control in all our data points.
In the following, all autocorrelation times are given in units of
molecular dynamics time.

\subsection{Tuning of the parameters\label{sec:detsim}}

The HMC algorithm has two tuning parameters, the trajectory length
$\tau_\mathrm{traj}$  and 
the accuracy with which the molecular dynamics equations are integrated.
With the field transformation, one also has to fix the integration
length $t_T$ of the field transformation and its step size.
Let us go through these parameters, the final values are listed in 
Table~\ref{tab:detsim}.

\begin{figure}[tb]
\begin{center}
\includegraphics[width=0.48\textwidth,clip]{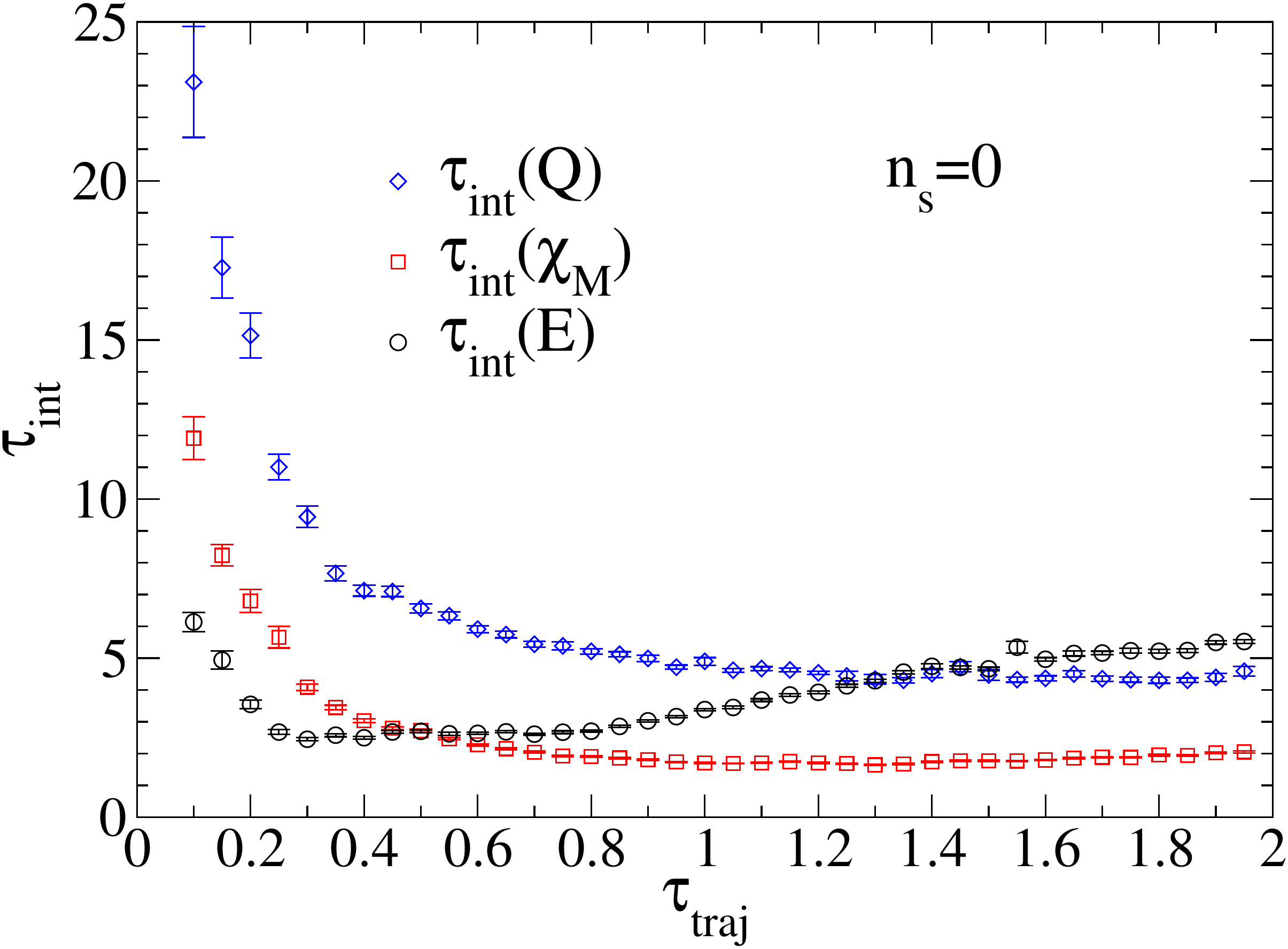}
\includegraphics[width=0.48\textwidth,clip]{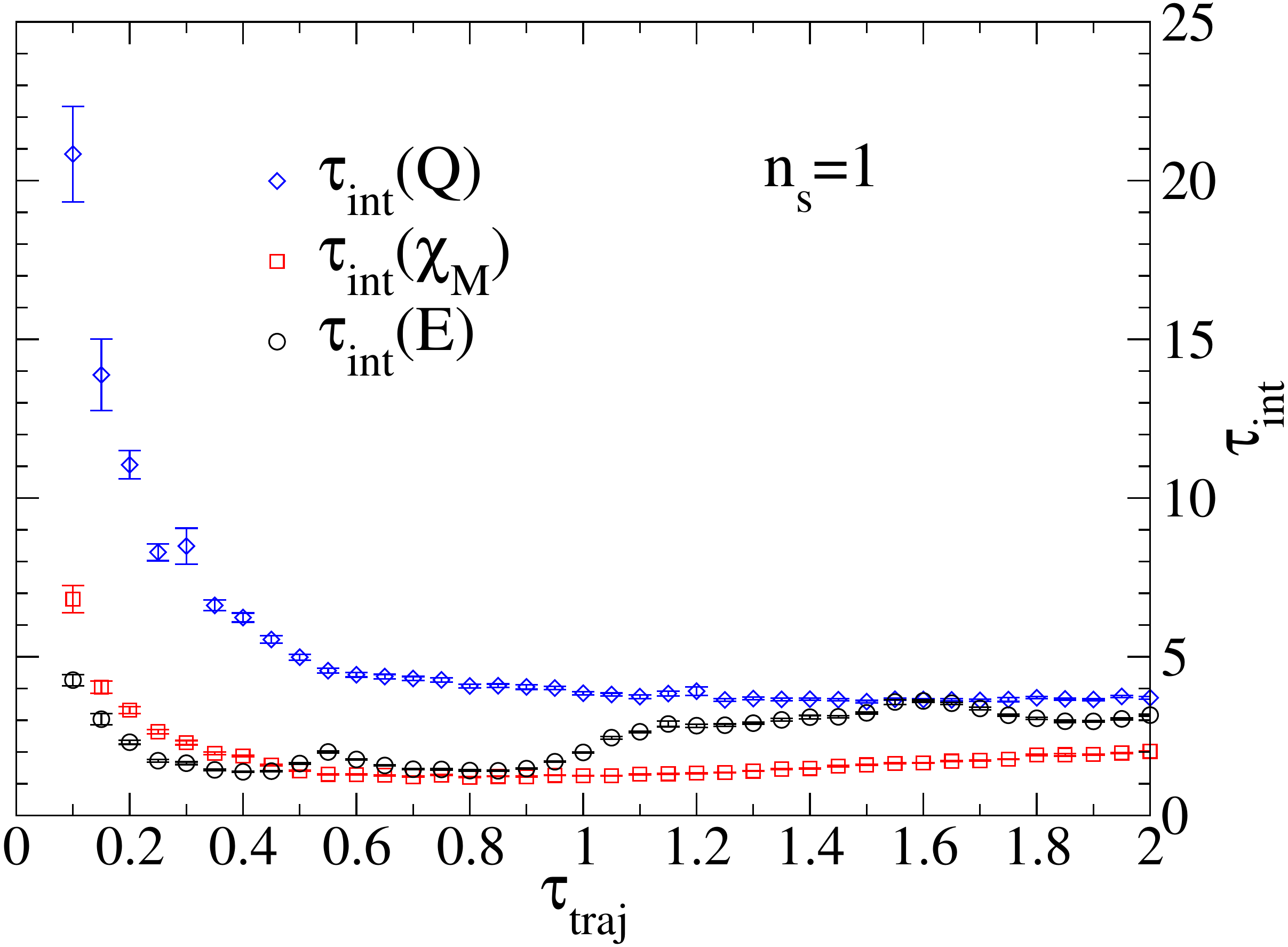}
\end{center}

\caption{Integrated autocorrelation time $\tau_\mathrm{int}$ versus the
trajectory length $\tau_\mathrm{traj}$ in CP$^9$, $\beta=0.7$.
$\tau_\mathrm{int}$ of the topological charge $Q$, the magnetic
susceptibility $\chi_M$ and the energy $E$ is  shown.  Left side HMC
without trivialization, right side THMC with one step of Euler
integration $n_s$ and integration of the flow to $t_T=0.47$, such that
the force is minimized.  For both HMC and THMC a choice of the
trajectory length around one is a good compromise.
\label{fig:1}
}
\end{figure}

\subsubsection{Trajectory length}
The effect of the trajectory length on the autocorrelation times  for
the CP$^9$ model with $\beta=0.7$ can be found in Fig.~\ref{fig:1}.  The
left hand plot shows the HMC without trivialization, the right hand side
the THMC with the leading order flow integrated up to $t_T=0.47$ (such
that the force is minimized, as will be discussed later) with one Euler
step ($n_s=1$).  The step size $\epsilon=\tau/n_\mathrm{step}$ of the
molecular dynamics trajectory integration is held approximately constant
in all data points. (In all runs we targeted acceptance rates
between 70\% and 90\%.)  As already expected from the QCD experience, the
optimal value of the trajectory length depends significantly on the
observable. The energy $E$ decorrelates fastest, with a clear minimum at
$\tau_\mathrm{traj}\approx0.3$, whereas the magnetic susceptibility
exhibits a very shallow minimum starting from
$\tau_\mathrm{traj}\approx1$. The topological charge $Q$ can profit from
even longer trajectories. This is the case for the standard HMC as well
as the one including the field transformation.  Considering the
computational costs of effectively decorrelated configurations,
$\tau_\mathrm{traj}$ in the range between 0.5 and 1 seems to be
efficient.  We therefore choose $\tau_\mathrm{traj}=1$ for all $\beta$
in our main runs as a compromise.  Although this might not be the
optimal choice, it is a standard practice in QCD simulations.

\begin{figure}[tb]
\begin{center}
\includegraphics[width=0.48\textwidth,clip]{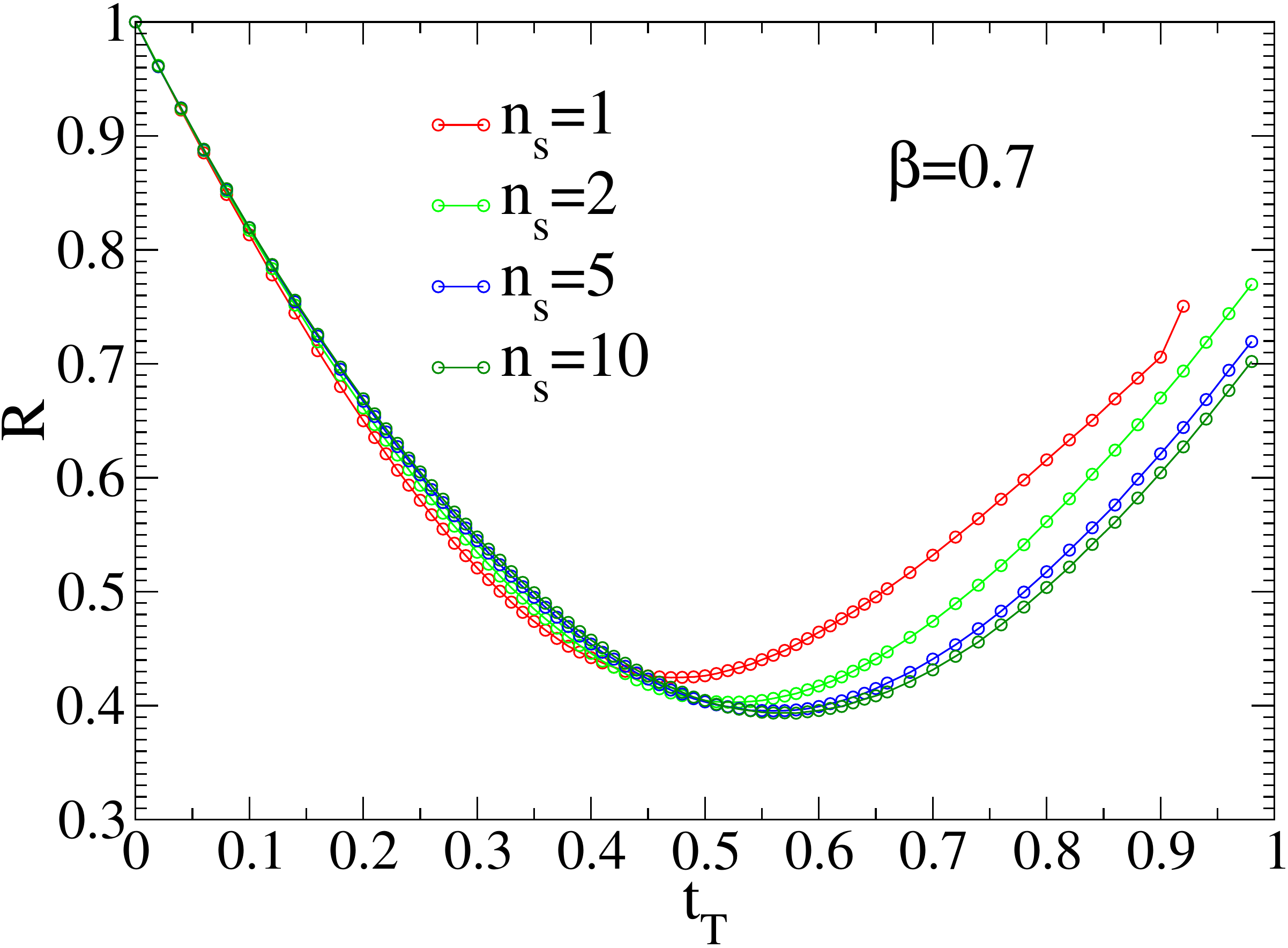}
\includegraphics[width=0.48\textwidth,clip]{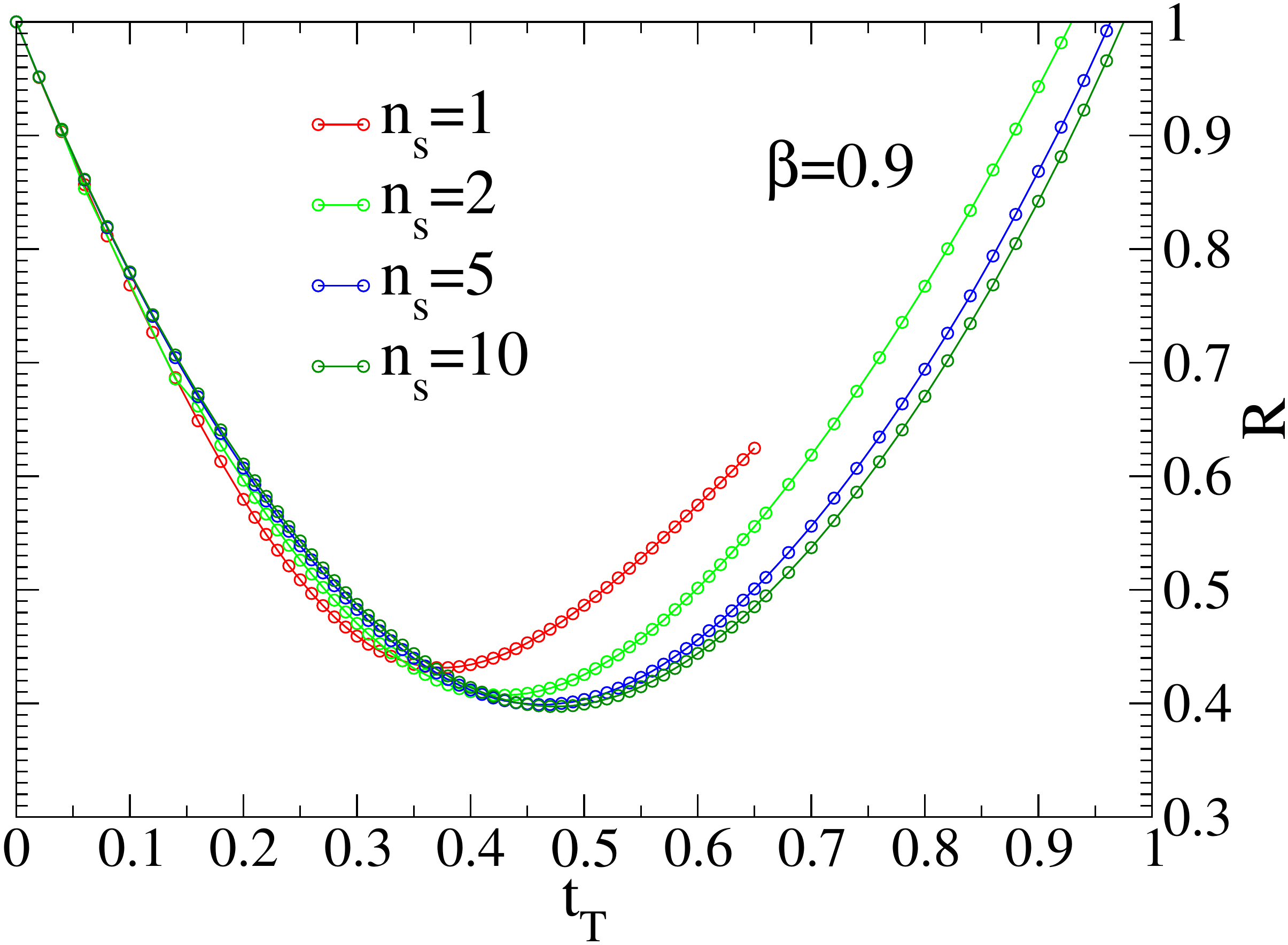}
\end{center}

\caption{\label{fig:2}Reduction of the force in THMC compared 
to the force in HMC ($R=F_{THMC}/F_{HMC}$) depending on the
integration length $t_T$ of the trivializing flow.  Data are shown for
CP$^9$ with 1,2,5 and 10 steps of the Euler integration.  Left side
$\beta=0.7$, right side $\beta=0.9$.  At large $t_T$, the curves are
ordered as in the legend, $n_s=1$ at the top and $n_s=10$ at the bottom.
The statistical errors are too small to be seen.  The improvement for
$n_s>1$ is negligible.} 
\end{figure}

\begin{figure}[tb]
\begin{center}
\includegraphics[width=0.55\textwidth,clip]{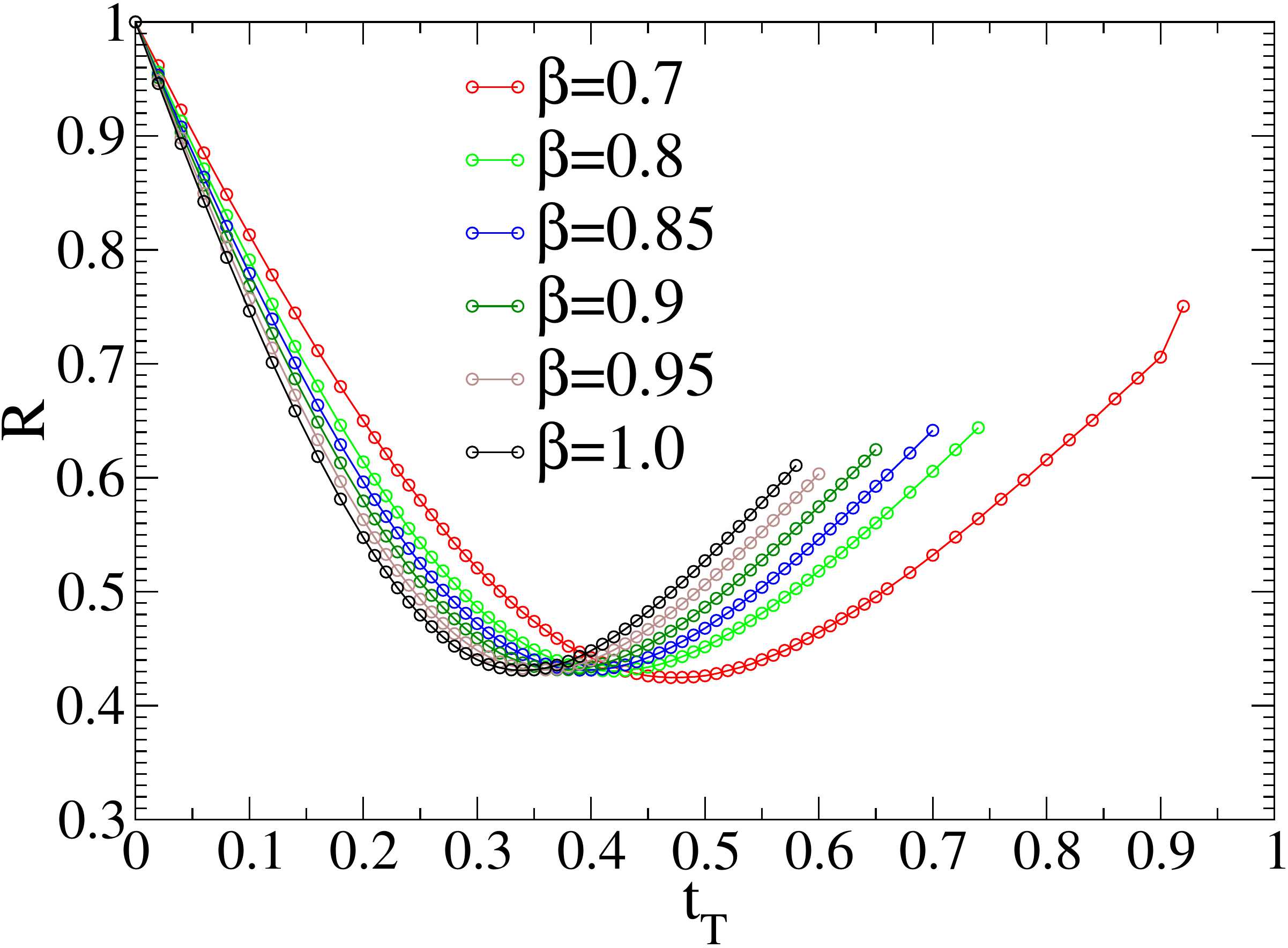}
\end{center}
\caption{Force reduction $R$ versus the flow integration length $t_T$, only 1 step of Euler integration ($n_s=1$), for several values of $\beta$ in CP$^{9}$.
At small $t_T$, the curves are ordered as in the legend, $\beta=0.7$ at the top and $\beta=1.0$ at the bottom.
As $\beta$ increases, the minimum moves towards smaller $t_T$, however, its depth does not change dramatically.
\label{fig:3}
}
\end{figure}

\subsubsection{Integration of the flow}
The second parameter to fix in our setup of the THMC is the value $t_T$
to which the flow Eq.~(\ref{eq:flow}) is integrated and the accuracy of
the integration, which is given by the number of steps in the Euler
integration.  As a criterion, we use the reduction of the forces
experienced in the molecular dynamics evolution, because perfect
trivialization would result in forces equal to zero.  The relative
reduction of the force $R=F_{THMC}/F_{HMC}$ for $\beta=0.7$ and
$\beta=0.9$ in the CP$^9$ model is shown in Fig.~\ref{fig:2}. In both
cases a reduction by about 60\% can be reached at a value of $t_T$
around 0.5, much smaller than the $t_T=1$ for which trivialization is
reached with the exact flow.  The force reduction depends very little on
the accuracy of the integration: whether 1, 2, 5 or 10 steps of the
Euler integrator are used hardly matters.  As shown in Fig.~\ref{fig:3},
the optimal value of $t_T$ decreases with increasing $\beta$, however,
the reduction of the force at the minimum is almost constant, at least
in the range $\beta=0.7,\dots,1.0$ which we have investigated.

\begin{figure}[tb]
\begin{center}
\includegraphics[width=0.48\textwidth,clip]{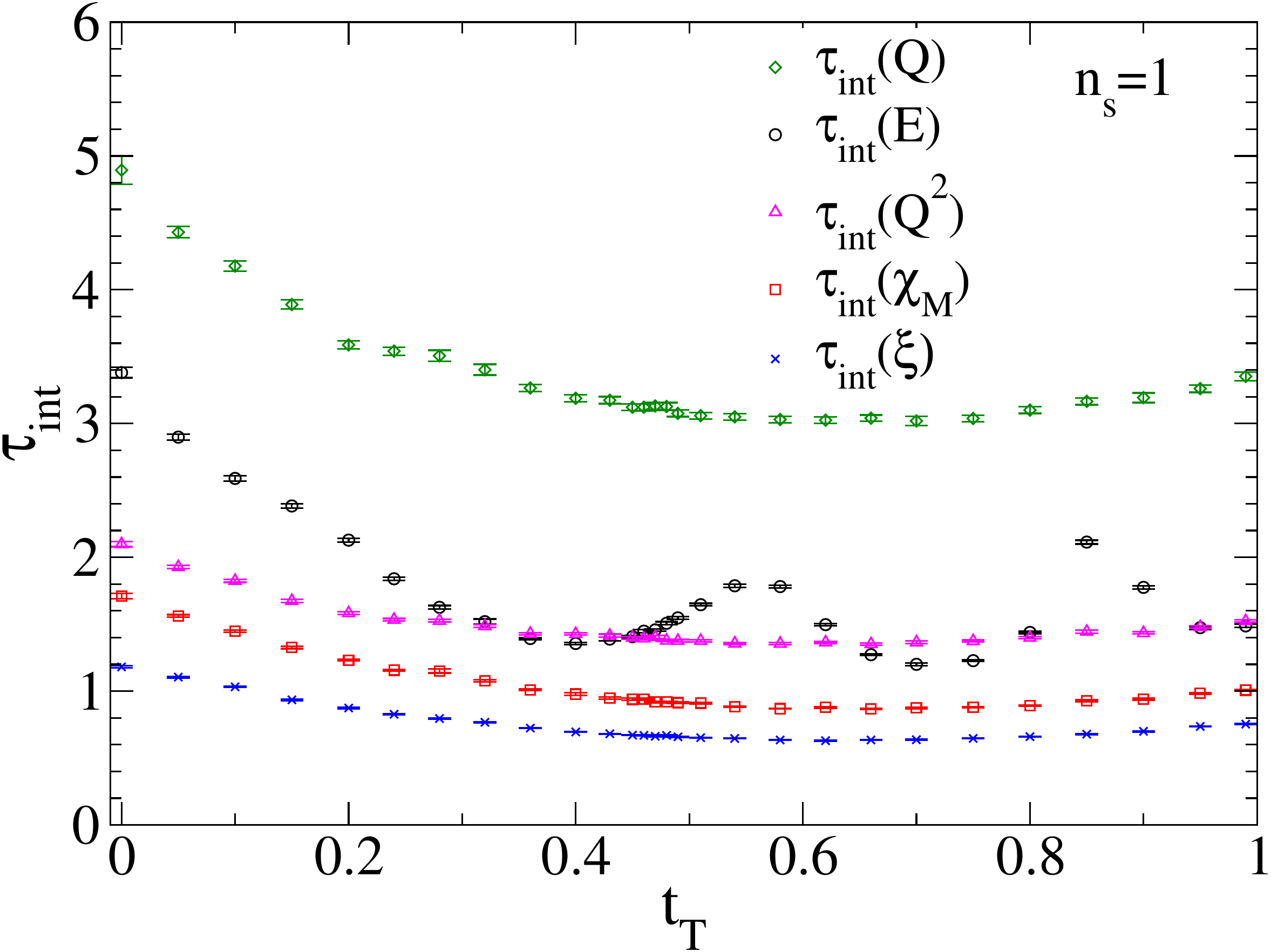}
\includegraphics[width=0.48\textwidth,clip]{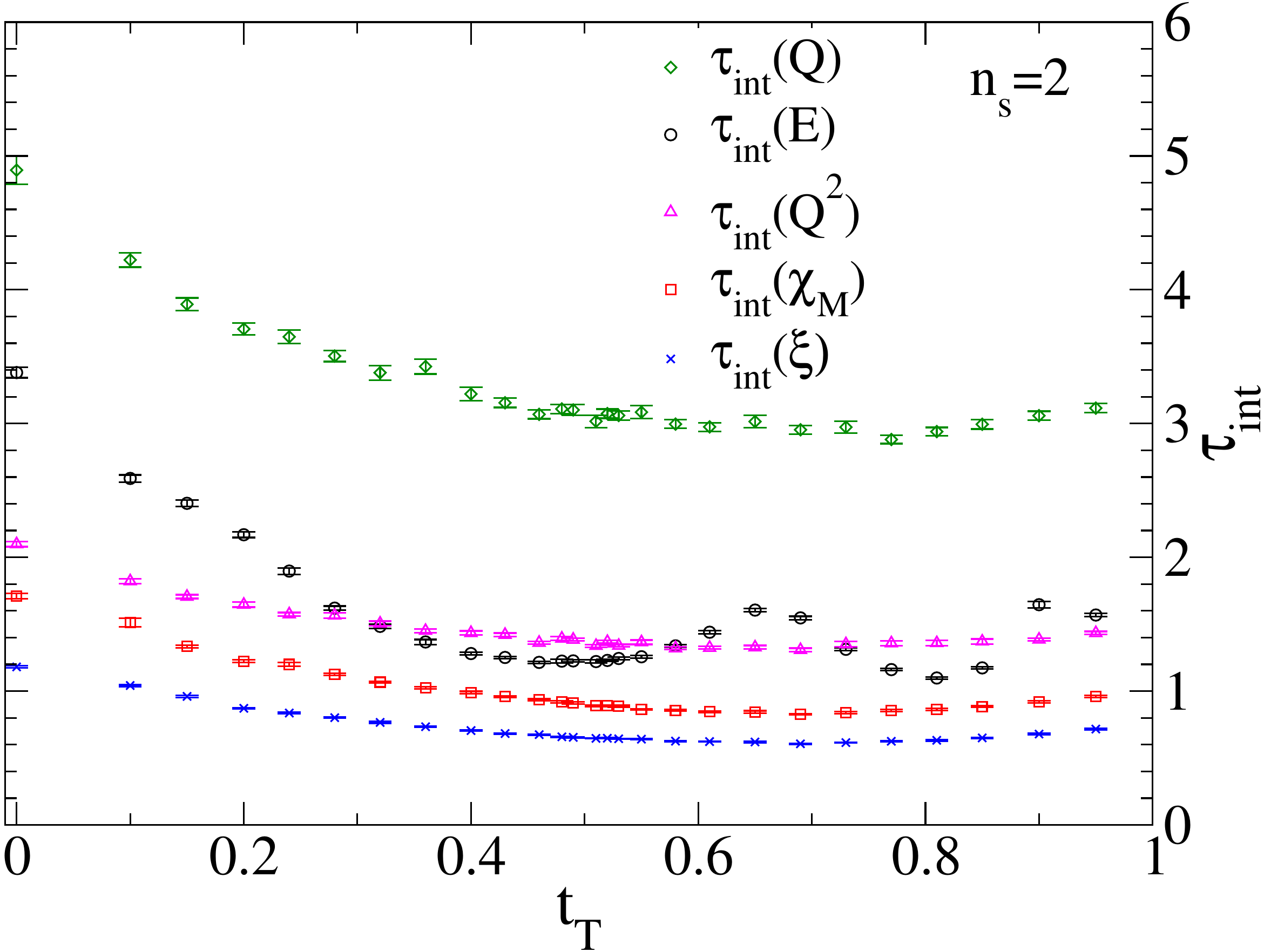}
\end{center}
\caption{Integrated autocorrelation time $\tau_\mathrm{int}$ versus the 
integration length $t_T$ of the leading order trivializing flow, 
for CP$^{9}$ with $\beta=0.7$. Left hand side: one step of Euler integration 
$n_s=1$. Right hand side: $n_s=2$. As shown in Fig.~\ref{fig:2}, maximum force 
reduction is obtained by $t_T=0.47$ ($n_s=1$) and $t_T=0.51$ ($n_s=2$), respectively.
\label{fig:4}
}
\end{figure}

Besides the reduction of forces, the field transformation is also
supposed to reduce the autocorrelations experienced in the
simulation. 
In Fig.~\ref{fig:4} we therefore show the integrated autocorrelation time for different
flow integration lengths $t_T$ for CP$^{9}$ with $n_s=1$ and $n_s=2$.
This can also be seen as a check of the force criterion depicted in
Figs.~\ref{fig:2} and \ref{fig:3}.  As expected, the optimum $t_T$ to
minimize autocorrelations depends on the observable considered.
Qualitatively, we observe that the force criterion leads to a good
choice of $t_T$ with respect to reduction of the autocorrelation.
However, while for larger values of $t_T$ the force ratio becomes worse,
the autocorrelation shows a fairly flat behavior. This test was done at fixed molecular dynamics step
size and therefore fixed cost per trajectory. The main reason for the
autocorrelation time rising for large $t_T$ is the decreasing 
acceptance rate. Had we kept the acceptance rate constant (meaning increased
costs for $t_T$ larger than the force minimum), the autocorrelation 
times would have decreased further, although not very much. 
We conclude that the force criterion is reasonable for tuning $t_T$
and we therefore used it throughout this study.

\subsubsection{Total cost of the simulation}
The reduction in the forces by about a factor two from the field
transformation allows a larger
molecular dynamics step size by about the same factor.
In particular for larger values of $\beta$, this leads to 
the same acceptance rate for HMC and THMC, see Table~\ref{tab:detsim}
for details.
In our implementation, an elementary leap-frog step of THMC with $n_s=1$ 
costs almost three times more than with HMC. More integration steps will 
further increase the cost. Together with the increased step size, this 
translates to roughly a factor of 1.5 increased cost per trajectory.
In the next section, we will find a reduction of the autocorrelation
times between roughly 1.5 and 1.8, depending on the observable. This
means that the total cost of simulation for HMC and THMC with $n_s=1$
are about the same.

\section{Results\label{sec:results}}

With this setup, we performed extensive runs at correlation lengths
between $\xi\approx2.3$ and $\xi\approx16.6$, using the plain HMC and
compare it to the THMC with the flow integrated with $n_s=1$ Euler step.
For the latter we use the optimal values
of the flow parameter $t_T$ with respect to reduction of the force.
The detailed parameters can be found in Table~\ref{tab:detsim},
expectation values of various observables in Table~\ref{tab:res}. The
measured autocorrelation times in units of molecular dynamics time are
listed in Table~\ref{tab:ac}.

\begin{table}
\begin{center}
\begin{tabular}{cccccccc}
 \hline
 \hline
$\beta$	& $L$	& $n_{s}$ &	$\tau_\mathrm{traj}$	&
$n_\mathrm{step}$   	&$t_T$		& $P_\mathrm{acc}$	&  stat.~[MD time] 	\\
 \hline
0.70    &   42  & 0   &  1       & \p62        &---      &0.84     &\p1000k        \\
0.70    &   42  & 1   &  1       & \p31        &0.47     &0.77     &\p5000k        \\
0.80    &   60  & 0   &  1       & \p85        &---      &0.85     &\p1000k        \\
0.80    &   60  & 1   &  1       & \p43        &0.43     &0.81     &\p4346k        \\
0.85    &   72  & 0   &  1       & \p97        &---      &0.86     &\p3682k        \\
0.85    &   72  & 1   &  1       & \p49        &0.40     &0.82     &\p4230k        \\
0.90    &   90  & 0   &  1       & 120       &---        &0.86     &\p2418k        \\
0.90    &   90  & 1   &  1       & \p60        &0.37     &0.85     &\p4114k        \\
0.95    &   120 & 0   &  1       & 170       &---        &0.90     &25282k       \\
0.95    &   120 & 1   &  1       & \p85        &0.35     &0.90     &13110k       \\
1.00    &   160 & 0   &  1       & 200       &---        &0.90     &32763k       \\
1.00    &   160 & 1   &  1       & 100       &0.33       &0.90     &14408k       \\
\hline
\hline
\end{tabular} 
\caption{Parameters of our runs in CP$^9$ with coupling $\beta$ and lattices of size $L^2$. 
$n_s$ is the number of Euler integration steps, where zero corresponds to the
standard HMC algorithm. 
$\tau_\mathrm{traj}$ denotes the integration length of the molecular dynamics
trajectory, $n_\mathrm{step}$ its discretization, 
$t_T$ the integration length of the trivializing flow and $P_\mathrm{acc}$ the Metropolis acceptance rate.
The last column gives the statistics in units of molecular dynamics time.
\label{tab:detsim}}
\end{center}
\end{table}

\begin{table}[tbp]
\begin{center}
\begin{tabular}{cccllll}
 \hline
 \hline
 $\beta$&  $L$  &$n_{s}$ & \p\p$\xi$     &  \p\p\p$E$             &\p\p\p$\chi_M$       &  $10^5 Q^2/V$  \\
 \hline
0.70    &   42  &0   &  \p2.312(3)   &  0.784378(16)    &\p10.124(3)    &  470.6(1.4)  \\
0.70    &   42  &1   &  \p2.3117(12) &  0.784361(6)     &\p10.1278(12)  &  470.6(6)   \\
0.80    &   60  &0   &  \p4.602(6)   &  0.667028(10)    &\p28.088(16)   & \p97.6(8)   \\
0.80    &   60  &1   &  \p4.595(2)   &  0.667023(4)     &\p28.068(6)    & \p96.9(3)   \\
0.85    &   72  &0   &  \p6.389(5)   &  0.622276(4)     &\p46.91(2)     & \p46.0(4)   \\
0.85    &   72  &1   &  \p6.386(4)   &  0.622271(4)     &\p46.916(14)   & \p46.2(3)   \\
0.90    &   90  &0   &  \p8.816(11)  &  0.583835(4)     &\p78.40(6)     & \p23.3(5)   \\
0.90    &   90  &1   &  \p8.837(6)   &  0.583834(3)     &\p78.49(3)     & \p23.3(3)   \\
0.95    &   120 &0   &  12.134(7)    &  0.5502611(8)    &131.39(5)      & \p11.73(16)   \\
0.95    &   120 &1   &  12.132(7)    &  0.5502626(11)   &131.41(5)      & \p11.91(19)   \\
1.00    &   160 &0   &  16.607(12)   &  0.5205860(6)    &220.48(12)     &\p\p6.18(18)  \\
1.00    &   160 &1   &  16.601(14)   &  0.5205872(8)    &220.37(13)     &\p\p6.14(20)  \\
 \hline
 \hline
\end{tabular} 
\caption{Expectation values of our runs in the CP$^9$ model, $n_s$ is the
number of Euler integration steps, further parameters are found in Table \ref{tab:detsim}. 
We give results for the correlation length $\xi$,
the energy, the magnetic susceptibility and the square of the
topological charge. \label{tab:res}}
\end{center}
\end{table}
\begin{table}
\begin{center}
\begin{tabular}{cccccccc}
 \hline
 \hline
 $\beta$& $n_{s}$ &  $\xi$&  $\tau(\xi)$   & $\tau(E)$  & $\tau(\chi_M)$ & $\tau(Q)$ & $\tau(Q^2)$\\
 \hline                                                                                                                      
   0.70    & 0   & 2.312(3)  &1.181(8) & 3.38(4)    &1.71(2)   &4.9(1)      &2.10(2)  \\
   0.70    & 1   & 2.3117(12)&0.943(3) & 1.981(8)   &1.268(8)  &3.86(2)     &1.792(7) \\
   0.80    & 0   & 4.602(6)  &3.61(6)  & 3.60(6)    &5.14(10)  & 35.3(1.2)  & 16.6(4)  \\
   0.80    & 1   & 4.595(2)  &1.983(13)& 2.99(4)    &2.84(2)   & 27.0(4)    & 12.30(13)\\
   0.85    & 0   & 6.389(5)  &7.32(9)  & 3.83(6)    &9.80(14)  &   126(4)   & 57.0(1.3) \\
   0.85    & 1   & 6.386(4)  &3.80(3)  & 3.62(5)    &5.29(6)   &  95(5)     & 43.7(8)  \\
   0.90    & 0   & 8.816(11) &13.8(3)    & 3.86(9)    & 18.4(5)&   527(38)  &  238(12)  \\
   0.90    & 1   & 8.837(6)  &7.57(12) & 3.73(4)    & 10.5(2)  &   345(17)  &  160(5)   \\
   0.95    & 0   &  12.134(7)  &27.9(5)    & 3.86(11)$_{-0.0}^{+0.3}$   & 40.5(8) & 2260(120)   &   1080(40) \\
   0.95    & 1   &  12.132(7)  &16.3(4)    & 3.60(7)$_{-0.0}^{+0.2}$    & 25.2(8) & 1630(70)    &  800(30)  \\
   1.00    & 0   &  16.607(12) &67(3)      & 5.6(4)$_{-0.0}^{+1.5}$     &  115(7) & 13400(1100) &   6300(400) \\
   1.00    & 1   &  16.601(14) &38(3)      & 4.3(2)$_{-0.0}^{+1.9}$     & 70(5)   & 9400(800)   &   3860(250) \\
\hline
\hline
\end{tabular} 
\end{center}
\caption{Auto-correlation times corresponding to
Table~\ref{tab:res}.
For $\beta=0.95$ and $\beta=1.0$, systematic errors for $\tau(E)$
account for the uncertainty in estimating the contribution
of the tail of the auto-correlation function.
\label{tab:ac}}
\end{table}

\subsection{Critical behavior}
This brings us to our main result, the critical slowing down of the
simulations as $\beta\to\infty$.  For large correlation length $\xi$,
the autocorrelation times are expected to grow as 
\begin{equation}
\tau_\mathrm{int}(A) \propto \xi^z
\label{eq:scaling}
\end{equation}
with $z$ the dynamical critical exponent. It depends, of course, on the 
observable $A$. This scaling is only expected for asymptotically large $\xi$, 
however, also with our limited range we can get an estimate of the
severeness of the problem and the reduction brought by the field
transformation.
Since we have periodic boundary conditions, topological sectors are
expected to form in the continuum limit. Because of the ensuing
barriers in the free energy,
Ref.~\cite{DelDebbio:2004xh} suggests for the topological charge an 
exponential behavior of the form
\begin{equation}
\tau_\mathrm{int}(Q^2)\propto \exp(c\, \xi^\theta) \ .
\label{eq:exp}
\end{equation}
However, as we will see below, the presence of such an exponentially
slow mode does also have an effect on all observables which 
do not completely decouple from it. 

\begin{figure}[tb]
\begin{center}
\includegraphics[width=0.45\textwidth,angle=-90,clip]{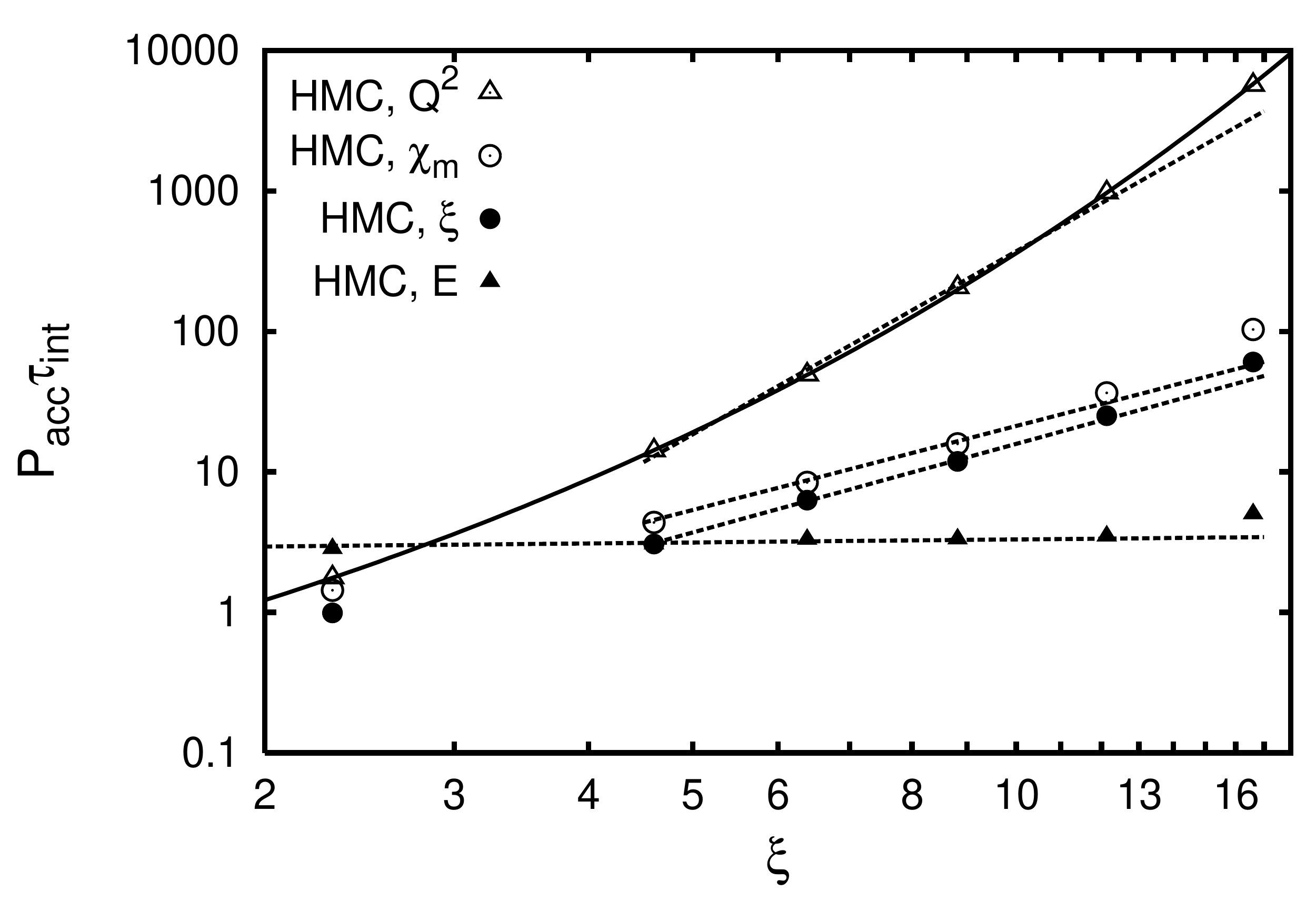}
\end{center}
\caption{The integrated autocorrelation time of various observables for
the HMC as a function of the correlation lengths. In this log-log plot, the dashed
lines indicate the results of  power law  fits to the data for
$4<\xi<13$.  The solid line represents the 
the exponential form Eq.~(\ref{eq:exp}). The error bars are smaller than
the size of the symbols.
\label{fig:5}
}
\end{figure}

In Fig.~\ref{fig:5} we show the $\tau_\mathrm{int}$ of various
observables as a function of the correlation length for the HMC
algorithm without field transformation. As expected, the slowing down in
the topological charge is much more severe than in the other observables
which in the interval $4<\xi<13$ show a behavior compatible with the
power law Eq.~(\ref{eq:scaling}). Making statistically relevant
statements about this is difficult, because the acceptance rates are not
constant over all our runs. We try to compensate for that by considering
$P_\mathrm{acc}\tau_\mathrm{int}$, but of course this is only a partial
correction. Also the data is not expected to follow exactly the leading
order scaling law; due to the high accuracy of our data
next-to-leading orders might become visible.  Nevertheless, fitting the
data in the range of $4<\xi<9$ to Eq.~(\ref{eq:scaling}), we get
$z=2.0(1)$ for the magnetic susceptibility and for the correlation
length.   The errors are statistical only.  The energy $E$ exhibits a
very flat behavior in $2<\xi<13$ with $z=0.12(1)$.  For $E$ and $\xi$,
the fits have a $\chi^2$/dof between 1 and 3.3, which is acceptable
considering the simple formula and the problems discussed above.
However, while the behavior for $E$ and $\xi$ are compatible with a
power law up to $\xi\approx12.1$, the last data point at $\xi\approx
16.6$, and for $\chi_m$ also the point at $\xi\approx12.1$, show a clear
deviation.  We interpret this as a consequence of a correlation between
these observables and the topological charge and will discuss this issue
below in detail. 

The square of the topological charge exhibits a much worse scaling
behavior than the other observables. Fitting Eq.~(\ref{eq:scaling}) to
the data with $4<\xi<12$ , we get $z\approx4$, however, the agreement is
not convincing and the $\chi^2$/dof$\;\approx20$ is poor. The
exponential function Eq.~(\ref{eq:exp}) works much better and delivers a
good description of the data in the whole region $2<\xi<17$ with
$c\approx4.2$ and $\theta\approx0.43$. It has
$\chi^2$/dof$\;\approx0.25$, but due to the problems discussed above,
this has to be taken with care.

\subsection{Effect of the slow modes}
Having detected at least one very slow mode in the simulation raises the
question to what extent the various observables are affected. The answer will
depend both on the particular observable and the accuracy required in the
simulation. We interpret the deviation from the power law scaling
behavior of the energy, the magnetic susceptibility and the correlation
length observed in Fig.~\ref{fig:5} at $\xi\approx16.6$ (and weakly already at $\xi\approx12.1$) to be a
consequence of the correlation between the slow mode and the observables.
As observed in the topological charge squared, the time constant of this mode
rises exponentially and at this point its contribution is no longer sufficiently
suppressed by the smallness of the coupling to the observable and
becomes noticeable. 

If we identify, for a moment, the slow mode with the topological charge,
one can understand the phenomenon with the following: while the
simulation is trapped in one topological sector, it samples the
observable restricted to that sector $A(Q)$. If the estimate obtained
before moving on to another sector after about $\tau(Q)$ steps is more
precise than variance $\Delta^2=\mathrm{var}_Q(A(Q))$ of $A(Q)$ over
topological sectors, then the autocorrelation time will have a
significant contribution from the topological modes. What matters thus
is the relative size of
$\sqrt{\mathrm{var}(A(Q))\tau_\mathrm{int}(A)/\tau_\mathrm{int}(Q^2)}$ and
$\Delta$. In our data we observe that while the former is larger than
the latter for most of our data points, this ordering is reversed for
the points at the largest correlation length.  If one is interested in
the level of accuracy given by $\Delta$, the simulation has to run over many
$\tau_\mathrm{int}(Q^2)$. 

\begin{figure}[tb]
\begin{center}
\includegraphics[width=0.45\textwidth,angle=-90,clip]{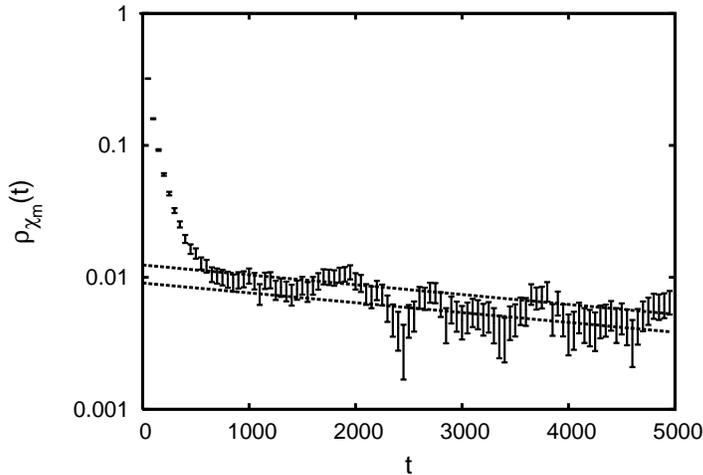}
\end{center}
\caption{Normalized autocorrelation function for the magnetic
susceptibility for the HMC algorithm at $\beta=1$. The dashed lines
correspond to a single exponential whose time constant has been
extracted from the autocorrelation function of $Q^2$. The coefficients
have been adjusted such that the two lines contain the $1\sigma$ region
around $t=1000$. This contribution accounts for roughly half the
integrated autocorrelation time.
\label{fig:6}
}
\end{figure}

As discussed in Ref.~\cite{Schaefer:2010hu}, the slow modes also pose a
problem for the accurate determination of the autocorrelation times
themselves. By restricting the sum in Eq.~(\ref{eq:tauint}) to some
window $W$, a small in amplitude but potentially long tail is neglected.
To illustrate this, we show the autocorrelation function of the magnetic
susceptibility  at $\beta=1$ for the HMC algorithm in Fig.~\ref{fig:6}.
At the beginning, it falls quickly to $\rho(t)\approx 0.01$ but then
develops a very long tail, a situation already described in
Ref.~\cite{Luscher:2010we}.
 The tail is compatible with a single
exponential with a time constant equal to
the exponential autocorrelation time extracted from
$\rho_{Q^2}(t)$. This is indicated in the figure by the dashed lines.
We can use this information for an improved estimate of the autocorrelation
time\cite{Schaefer:2010hu}. The usual sum of
$\rho(t)$ in only performed up to the point where the single exponential
tail starts. The rest of the sum is substituted by  the integral over 
the single exponential for which the largest observed time constant
observed in all observables with the same parity is taken. In our
situation this is $\tau_\mathrm{exp}(Q^2)$. Since the 
HMC obeys detailed balance, this gives
a strict upper bound for $\tau_\mathrm{int}$, provided that there are
no modes which suffer from an even slower evolution.

Also for the correlation length and  at $\beta=0.95$ a similar behavior
can be observed. Even though the coefficients might seem small, the slow
modes still have a sizable contribution because of  the very
large time constant.  At $\beta=1$, this tail contributes roughly
30\% to $\tau_\mathrm{int}(\xi)$ and 50\% to
$\tau_\mathrm{int}(\chi_m)$; for $\beta=0.95$ the contribution of
the tail is roughly 10\% and 17\%, respectively.  The values of the
autocorrelation time from estimating the contribution of the tail in
this way lie within the $1\sigma$ error of the values given
in Table~\ref{tab:ac} obtained by a summation to large values of  $W$.
In case of $\tau_\mathrm{int}(E)$, the improved estimator is
significantly higher than the value obtained from the truncated sum.
The single exponential dominates from $t\approx150$, contributing
roughly 10\% (50\%) at $\beta=0.95$ ($\beta=1$).  We take the
different values as upper and lower bounds and state the discrepancy as
systematic error in Table~\ref{tab:ac}.  If the true value is close to
the upper bound, the scaling of $\tau_\mathrm{int}(E)$ deviates from a
power law already at $\xi\approx12.1$.  If we assume that the
exponential growth in the time constant is not compensated by the
decrease in the coefficient, this contribution will be even more
pronounced when $\beta$ is increased further.

\subsection{Performance of the field transformation}
\begin{figure}[tb]
\begin{center}
\includegraphics[width=0.45\textwidth,angle=-90,clip]{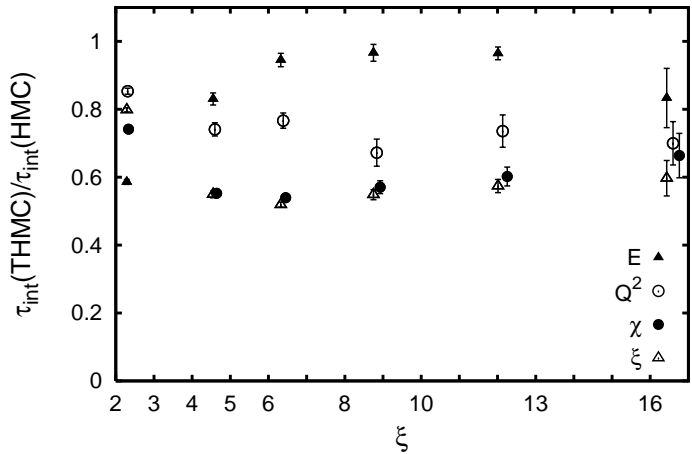}
\end{center}
\caption{Cost reduction in terms of autocorrelation time for the
different observables brought by the field transformation.
\label{fig:7}
}
\end{figure}

We finally come to the comparison between the HMC and THMC algorithm,
and we show the reduction in autocorrelation time achieved through the
introduction of the field transformation.  In Fig.~\ref{fig:7} we plot
the ratio of the  autocorrelation of our observables for the two
algorithms. The correlation length and the magnetic susceptibility for
which we observe a 40\% reduction profit most from the field
transformation. For the topology roughly a 25\% reduction is found, the
energy is almost unaffected, however, it shows a quite short $\tau_\mathrm{int}$ over the whole
range of data.
Note that for $\xi\approx12.1$ and $\xi\approx 16.6$ the reduction of 
$\tau_\mathrm{int}(E)$ has to be taken with care due to its systematic uncertainty.
All critical exponents of the THMC algorithm extracted
from the range $4<\xi<13$ agree within uncertainties with the ones of HMC.
Also the deviation from the scaling law at $\xi\approx12.1$ and $\xi\approx 16.6$ is
observed.  The exponential behavior of $\tau_\mathrm{int}(Q^2)$ is
compatible with HMC within error bars as well.  We can conclude that the
field transformation does not affect the scaling of these variables in
the investigated region.

As commented above, the improvement factor which we find is close to the
additional cost of the simulation and therefore the two algorithms
perform rather similarly. These findings do not seem to depend strongly
on $N$, since we also did limited simulations in the CP$^{20}$, with
essentially equal results.

\section{Summary}
Whether a modification to an algorithm actually improves its performance
is often very difficult to predict. It therefore needs numerical
simulations to study its effects. Here we investigated
recently proposed field transformations which can lead to a speed-up
in HMC simulations. Unfortunately, the result is negative. Although we
observe a reduction in autocorrelation times, the scaling towards the
continuum limit is not improved. The reduction in the forces,
which can be used to increase the step size of the molecular dynamics
integration, is compensated by the computational overhead of the method.
However, this conclusion does not have to be universally true for all
theories. In QCD with dynamical fermions, e.g., the computational cost
of the construction would be a minor part of the whole cost of the
simulation.

Investigating the pure HMC algorithm serves also as an illustration that
exponentially slow modes will at some point affect other observables of
the theory.  The deviation at $\xi\approx16$ from the scaling behavior,
which we observe for the observables up to a correlation length of 12, is
therefore a cautionary tale for QCD simulations. Even though the slow
mode observed in the topological charge  might not seem to
have any influence on other observables in today's 
simulations\cite{Schaefer:2010hu}, at some
point, the correlation to the topological charge can also affect the 
scaling behavior towards the continuum limit in these channels.

\subsection*{Acknowledgements}
The calculations have been performed on local clusters at ZID at the
University of Graz. We thank the institution for providing support.  
We are indebted to Christof Gattringer and Martin L\"uscher for
very valuable comments on a previous version of the paper and to
Francesco Virotta for providing us with software for the
analysis of autocorrelation functions.  G.P.E.~is  grateful for the hospitality 
of the Computational Physics group of the Humboldt Universit\"at zu Berlin 
and acknowledges support by the Doktoratskolleg DK W1203-N08 and the DFG project SFB/TR-55.

\providecommand{\href}[2]{#2}\begingroup\raggedright\endgroup

\end{document}